\documentclass[prl,twocolumn,showpacs,amsmath,amssymb,superscriptaddress, nobibnotes]{revtex4-1}
\pdfoutput=1

\usepackage[latin1]{inputenc}
\usepackage[dvipsnames]{xcolor}
\definecolor{KBFIred}{RGB}{163,35,47}
\usepackage{mathtools}
\usepackage{lipsum}

\usepackage{hyperref}

\newcommand{\nn}{\nonumber}

\newcommand{\pd}{\partial}

\newcommand{\hc}{\text{ H.c. }}

\newcommand{\KBFI}{Laboratory of High Energy and Computational Physics, National Institute of Chemical Physics and Biophysics, R\"avala pst.~10, 10143 Tallinn, Estonia}

\providecommand{\f}[2]{\frac{{#1}}{{#2}}}
\newcommand{\be}{\begin{equation}} 
\newcommand{\ee}{\end{equation}}

\def\baq{\begin{eqnarray}}
\def\eaq{\end{eqnarray}}

\newcommand{\beq}{\begin{equation}} 
\newcommand{\eeq}{\end{equation}}

\newcommand{\ie}{\emph{i.e.}}

\newcommand{\Eq}[1]{Eq.~(\ref{#1})}

\newcommand{\gsim}{\lower.7ex\hbox{$\;\stackrel{\textstyle>}{\sim}\;$}}
\newcommand{\lsim}{\lower.7ex\hbox{$\;\stackrel{\textstyle<}{\sim}\;$}}

\begin{document}

\title{Higgs-like spectator field as the origin of structure}

\author{Alexandros Karam}
\email{alexandros.karam@kbfi.ee}
\affiliation{\KBFI}

\author{Tommi Markkanen}
\email{tommi.markkanen@kbfi.ee}
\affiliation{\KBFI}
\affiliation{Helsinki Institute of Physics, P.O. Box 64, FIN-00014 University of Helsinki, Finland}

\author{Luca Marzola}
\email{luca.marzola@cern.ch}
\affiliation{\KBFI}

\author{Sami Nurmi}
\email{sami.t.nurmi@jyu.fi}
\affiliation{Helsinki Institute of Physics, P.O. Box 64, FIN-00014 University of Helsinki, Finland}
\affiliation{Department of Physics, University of Jyv\"askyl\"a, P.O. Box 35, FI-40014 University of Jyv\"askyl\"a, Finland}

\author{Martti Raidal}
\email{martti.raidal@cern.ch}
\affiliation{\KBFI}

\author{Arttu Rajantie}
\email{a.rajantie@imperial.ac.uk}
\affiliation{Department of Physics, Imperial College London, London, SW7 2AZ, United Kingdom}

\begin{abstract}
\noindent 

We show that the observed primordial perturbations can be entirely sourced by a light spectator scalar field with a quartic potential, akin to the Higgs boson, provided that the field is sufficiently displaced from vacuum during inflation. 
The framework relies on the indirect modulation of reheating, which is implemented without any direct coupling between the spectator field and the inflaton and does not require non-renormalisable interactions. The scenario gives rise to local non-Gaussianity with $f_{\rm NL}\simeq 5$ as the typical signal. As an example model where the indirect modulation mechanism is realised for the Higgs boson, we study the Standard Model extended with right-handed neutrinos. For the Standard Model running we find, however, that the scenario analysed does not seem to produce the observed perturbation.

\end{abstract}

\maketitle

\noindent\textbf{Introduction.}
The cosmological perturbations observed in the Cosmic Microwave Background (CMB) are most commonly interpreted as a product of the inflaton field quantum fluctuations, which are stretched to macroscopic scales during the inflationary expansion in the early Universe. 

Alternatively, the observed perturbation could be sourced by fluctuations of spectator fields which have no dynamical effect during inflation but affect the dynamics later on. Well-known examples include the curvaton model~\cite{Lyth:2001nq, Moroi:2001ct, Enqvist:2001zp, Linde:1996gt, Mollerach:1989hu} and the modulated reheating scenario~\cite{Kofman:2003nx, Dvali:2003em}. In the latter, the inflaton field decay width is modulated by spectator field fluctuations which sources the curvature perturbation and typically gives rise to significant non-Gaussianities~\cite{Ichikawa:2008ne, Litsa:2020mvj}. Modulated reheating has been extensively studied in the past, most usually in setups which rely on the direct coupling of the inflaton field to light spectator scalars -- for instance, the Higgs boson. We refer the reader to Ref.~\cite{Ichikawa:2008ne} for the general formalism of modulated reheating, and to Refs.~\cite{Kobayashi:2011hp, Fujita:2016vfj, Lu:2019tjj, Choi:2012cp, DeSimone:2012gq, Cai:2013caa,  Freese:2017ace, Lu:2019tjj, Litsa:2020rsm, Litsa:2020mvj} for models with a specific focus on the Higgs field. Further related work can be found in Refs.~\cite{Chambers:2007se, Chambers:2008gu, Chambers:2009ki, Fujita:2013bka, Fujita:2014hha}. 
Spectator field fluctuations can also indirectly modulate the reheating, by introducing spatial dependence in the inflaton decay width through kinematic blocking and thereby source the primordial perturbation ~\cite{Fujita:2016vfj, Lu:2019tjj, Choi:2012cp, Cai:2013caa,  Freese:2017ace, Karam:2020skk}. In our previous work~\cite{Karam:2020skk}, we analysed the consequences of the Standard Model (SM) Higgs field in this framework, assuming that prior to the reheating process the field was in the vacuum state determined by the equilibrium configuration on a de Sitter background~\cite{Starobinsky:1986fx, Starobinsky:1994bd}. Although the Higgs boson does give rise to a significant scalar perturbation spectrum, such equilibrium state generically leads to power spectra characterised by a blue tilt~\cite{Markkanen:2019kpv, Herranen:2013raa}, incompatible with the observed red-tilted CMB spectrum. In spite of that, quite naturally, spectator fields not in their vacuum state can generate the observed spectrum. Both in the modulated reheating and curvaton models, primordial perturbations with a red-tilted spectrum can be generated by spectator fields that possess non-vanishing mean values and slowly roll towards their respective vacuum states. In the curvaton setup, a crucial requirement is that after the end of inflation the spectator energy density needs to grow comparable to the dominant energy component in order for the mechanism to source significant perturbations. In the case of the Higgs boson, or of any spectator field with a quartic potential, this is problematic because the energy density of the field never dilutes slower than the dominant radiation component during the Hot Big Bang epoch~\cite{Figueroa:2016dsc, Choi:2012cp}. On the other hand, in the modulated reheating scenario, such constraint can be avoided as the efficiency of the mechanism does not necessarily depend on the spectator energy density.  

In this work we study a modulated reheating mechanism that allows a Higgs-like spectator field $h$ -- that is, driven by a quartic potential -- to source the observed primordial perturbation spectrum despite having a subdominant energy density throughout the cosmic evolution. The setup is similar to that of Ref.~\cite{Markkanen:2019kpv}, but in this work we concentrate on the limit of large spectator field values which allows to generate the observed red spectral tilt. We discuss how the mechanism could be implemented in extensions of the SM, with the purpose of identifying the spectator field with the Higgs boson. The main conclusion of our work is that a set-up with a quartic spectator with no coupling to the inflaton and without non-renormalizable operators can solely generate a curvature perturbation that is consistent with current observational bounds. 

\noindent\textbf{The model.}
Following Ref.~\cite{Karam:2020skk}, we consider a Lagrangian given by 
\begin{align}
	\label{eq:lag2}
	{\cal L} &= \frac{1}{2}(\pd\phi)^2 -\frac{1}{2}m^2_\phi\phi^2-\frac{\lambda_{\phi}}{4}\phi^4  + \frac{1}{2}(\pd h)^2 -\frac{\lambda_h}{4} h^4  \\&+ i \bar\Psi{\pd}\!\!\!/ \Psi \nn-y_\phi\bar\Psi\Psi\phi - y_h\bar\Psi\Psi h +\hc,
\end{align}
where $h$ is the spectator scalar field, the scalar singlet $\phi$ is the inflaton and $\Psi$ is a Dirac fermion. The form of the inflaton potential considered in Eq.~(\ref{eq:lag2}) is not a prerequisite of our setup and is only used to provide a simple and concrete template. 

It is crucial to our purposes that the decay rate of the inflaton is subject to the indirect modulation from a spectator field,
\begin{equation}
	\Gamma(h)=\frac{y_\phi^2  m_\phi}{8\pi}\left[1-\frac{(2y_hh)^2}{  m_\phi^2}\right]^{3/2}\,,\label{eq:G}
\end{equation}
such that the decay channel is blocked by kinematics when the spectator field has the value
\be 
\label{hkindef}
h\geqslant \f{m_\phi}{2y_h}\equiv h_{\rm kin}\,.
\ee
Since the Lagrangian also allows for decay of the spectator field, for simplicity we restrict here to parameters such that the effective mass of the spectator field, $\sqrt{3\lambda}h$, is much smaller than that of the fermion, $y_h h$, and thus the process can be safely ignored. It is worth pointing out that if we identify $h$ with the SM Higgs boson, the remaining couplings to SM fields, such as gauge bosons, can be safely neglected as they do not lead to any significant depletion of the Higgs field value over time scales relevant for the modulation effect,  corresponding to ${\cal O}(1)$ Higgs oscillations after the inflationary expansion~\cite{Enqvist:2015sua}. Note also that while the tree-level action (\ref{eq:lag2}) does not directly couple the inflaton to the Higgs boson, such a coupling is generated by fermions at the loop level. However, because of the small values of the Yukawa coupling $y_h$ that we are considering in this work, we neglect these radiative effects.

Denoting by $h_*$ the field value at the horizon crossing of a mode $k_{*} = a_* H_*$ during inflation, we now focus on the case $h_* > h_{\rm kin}$, indicating that the inflaton decay is kinematically blocked by the spectator field via the indirect modulation mechanism. We assume the $h$-field to be light if compared to the inflation scale, hence it  fluctuates \textit{locally} around its VEV and the kinematic blocking of the inflaton decay is thus lifted at different times in different locations in the Universe. This implements the modulated reheating mechanism~\cite{Dvali:2003em} in our scheme and allows for the production of significant curvature perturbations~\cite{Karam:2020skk, Ichikawa:2008ne,Choi:2012cp}. Specifically, we require that the spectator field effective mass satisfies $V''(h_*)/(3 H_*^2) < 0.01 $, which for our model translates to $\lambda_h h_*^2 < 0.01 H_*^2$. For $h_* < M_{\rm P}$, the condition implies  $\Omega_{h*} \lesssim 10^{-3}$, so the spectator field energy density is necessarily subdominant with respect to the inflaton contribution. 

As for the spectrum, at the equilibrium, spectator field fluctuations in a de~Sitter background result only in a blue tilted spectra~\cite{Markkanen:2019kpv, Herranen:2013raa} ($n_s > 1$) and therefore cannot source the observed primordial perturbation. The situation is different for field configurations far from equilibrium, \ie, in the mean field limit $h_* > \sqrt{\langle h^2\rangle}_{\rm eq},$  when fluctuations around the mean field value $h_*$ yield either red or blue tilts. 
In the setup specified by Eq.~(\ref{eq:lag2}), the out-of-equilibrium configuration $h_* > \sqrt{\langle h^2\rangle}_{\rm eq}$ corresponds to an atypical field configuration, or initial condition, which we suppose to be accidentally realised in our observable patch. This is similar to what is often assumed in the context of the curvaton scenario, for example see~\cite{Lyth:2001nq}. 
Alternatively, one might think of modifying the spectator field potential such that the initial condition could be dynamically realised.  Further study of this question is however beyond the scope of this work.   Here we simply assume the mean field limit $h_* > \sqrt{\langle h^2\rangle}_{\rm eq}$ due to the phenomenological reason that it is required to get the observed spectral tilt.
Using the de~Sitter equilibrium result $\sqrt{\langle h^2\rangle}_{\rm eq}\simeq 0.36 H_*/\lambda_h^{1/4} $~\cite{Starobinsky:1994bd}, the validity of the mean field limit requires $h_* > 0.36 H_*/\lambda_h^{1/4}$. Combining this with the mass bound $\lambda_h h_*^2 < 0.01 H_*^2$, our constraints are specified by 
\beq
\label{meanfield}
\frac{0.36}{\lambda_h^{1/4}} < \frac{h_{*}}{H_*}< \frac{0.1}{\lambda_h^{1/2}}~,
\eeq
which further implies the bound $\lambda_h < 0.006$. 

\noindent\textbf{Analytical estimates.} Simple analytical estimates of the spectrum and non-Gaussianity of curvature perturbations produced by the indirect modulation mechanism can be obtained in the limit where the inflaton decays rapidly after the kinematic blocking is lifted. This occurs when $h(t_{\rm kin}) = h_{\rm kin}$ for the first time in the evolution of the spectator field value. In the discussion, we neglect the slow evolution of $h$ during the inflationary expansion and approximate its value at the end of this process by $h_{\rm end} \simeq h_{*}$. The spectator field starts to oscillate when $H_{\rm osc} \approx \sqrt{3\lambda_h} h_*,$ and the field behaviour prior to the first zero crossing can be approximated quite accurately by~\cite{Karam:2020skk}
\beq
\label{eq:hevolution}
h\approx h_*\left(1-\frac{3}{2}{\rm e}^{-\frac{27}{4}\frac{H}{\sqrt{3\lambda_h} h_*}}\right)\,,
\eeq
where $H \propto a^{-3/2}$ for a Universe dominated by inflaton oscillations in a quadratic potential. 

In order to ensure the prompt decay of the inflaton field during the first available window allowed by kinematics, we impose that $\rho_{\phi}/\rho_{\rm tot}$ falls below $10^{-5}$ as $|h| < h_{\rm kin}$ during the first half of the first $h$ oscillation -- i.e. the first period when $h$ moves from $h_{\rm kin}$ through zero to $-h_{\rm kin}$. For $\rho_{\phi}/\rho_{\rm tot} < 10^{-5}$, the inflaton contribution to the number of $e$-folds from later times, ${\cal O}(\rho_{\phi}/\rho_{\rm tot})\int H dt$, can be neglected when studying perturbations of order $\Delta N \sim 10^{-5}$. As a first approximation, we can then approximate that the Universe changes from matter to radiation domination at $t=t_{\rm kin}$, and from Eq.~\eqref{eq:hevolution} it follows that
\begin{equation}
\label{Hdec}
	H_{\rm kin}(h_*) = 
	\frac{4}{27} \sqrt{3\lambda_h}h_*\, \ln\left(\frac{3  h_*}{2(h_* - h_{\rm kin})}\right)\,.
\end{equation}
By using $\rho_{\phi}\propto {\rm exp}(-\int\Gamma dt)$, and estimating the integral over the first window allowed by kinematics as $\int \Gamma dt \sim 2 \Gamma_0 \int_{t_0}^{t_k} dt$, where $t_0$ and $t_k$ are determined by solving Eq. (\ref{eq:hevolution}) for $h(t_k) = h_{\rm kin}$ and $h(t_0) = 0$, the fast decay condition $-\int \Gamma dt > {\rm ln} 10^{-5} $ translates into the bound
\beq 
\label{eq:modc}
h_* \lesssim 0.21 m_{\phi} y_{\phi}y_h^{-1/2} \lambda_h^{-1/4}~.
\eeq
Here we have also assumed the limit $h_{\rm kin}\ll h_*$. The numerical analysis presented below indeed confirms that this inequality well determines the regime where our analytical approximations are fully applicable. 

The curvature perturbation in Fourier space can be computed in terms of the $\delta N$ expression~\cite{Wands:2000dp} as
\beq
\label{eq:deltaN}
\zeta({\bf k_*}) = N'\delta h_*({\bf k_*}) + \frac{1}{2} N''\delta h_*({\bf k_*})^2+ \ldots ,
\eeq
where $N$ is the number of $e$-folds from the horizon crossing $k_* = a_*H_*$ to a final constant energy surface after the inflaton decay, a prime denotes differentiation with respect to the field value $h_*$ at horizon crossing, and $\delta h_*^n$ are convolutions. Throughout this work we concentrate on the limit where the inflaton contribution to the curvature is negligible, $\zeta_{\phi} \approx (1/\sqrt{2\epsilon_*}) H_*/(2\pi M_{\rm P}) \ll 10^{-5}$, and therefore omit it in all expressions. 

When the condition in Eq.~\eqref{eq:modc} is satisfied, we can well approximate the evolution of the Universe using a sudden transition from matter to radiation domination at $t_{\rm kin}$, which yields~\cite{Karam:2020skk} $N' \approx - (1/6) {H_{\rm kin}'}/{H_{\rm kin}}$, with $H_{\rm kin}$ given by Eq.~\eqref{Hdec}. Substituting this into Eq.~\eqref{eq:deltaN}, the power spectrum of curvature perturbation becomes
\begin{align}
\label{eq:Pana}
{\cal P}_\zeta(k_*)=\f{H'_{\rm kin}{}^2}{36 H_{\rm kin}^2}
\left(\frac{H_*}{2\pi}\right)^2
\approx \f{1}{36 h_{*}^2}\left(\f{H_{*}}{2\pi}\right)^2
+ {\cal O} \left(\frac{h_{\rm kin}}{h_*}\right)^2,
\end{align}
where in the last step we expanded Eq.~\eqref{Hdec} in the limit $h_{*}\gg h_{\rm kin}$. The spectral index is given by~\cite{Wands:2002bn}
\be
\label{eq:nsana}
n_s-1=-2\epsilon_*+2\f{\lambda_h h_{*}^2}{H_*^2}~,
\ee
where $\epsilon\equiv-\dot{H}/{H^2}$ is computed at the horizon crossing. 

Comparing our result with the observations, ${\cal P}_{\zeta}(k_{*})=(2.100\pm0.030)\times 10^{-9}$ and $n_{\rm s}(k_{*}) = 0.965\pm 0.004$ at the pivot scale $k_{*} = 0.05\ {\rm Mpc}^{-1}$~\cite{Aghanim:2018eyx}, we find that the analytical estimates agree with the measurements for $h_*/H_* \simeq 580$ and $\epsilon_{*}\simeq 0.018$ 
(assuming $\lambda_h \lesssim 10^{-8}$ so that the positive contribution from the second term in Eq.~\eqref{eq:nsana} is negligible).  
For $h_* \simeq 580 H_*$, the assumed mean field condition~(\ref{meanfield}) implies $\lambda_h \gtrsim 1.5\times 10^{-13}$. Note also that the derivation of the estimate~(\ref{eq:Pana}) assumes the fast decay condition in Eq.~\eqref{eq:modc}, which constrains from above the range of $\lambda_h$ values for which the analytical estimate can be used. For instance, taking $y_{\phi} = 1, y_{h} =10^{-3}, m_{\phi}/H_* = 0.1$ and setting $h_* = 580 H_*$, Eq.~\eqref{eq:modc} yields $\lambda_h \lesssim 1.7 \times 10^{-12}$ and, for this particular example, cases with larger values of $\lambda_h$ need to be studied numerically -- see the results further below.

The second order term in Eq.~\eqref{eq:deltaN} contributes to the local bispectrum with a non-Gaussianity amplitude $f_{\rm NL} = (5/6) N''/(N'{}^2)$ given by 
\be
\label{eq:fnlana}
f_{\rm NL} = 5\left(1-\frac{H_{\rm kin}''H_{\rm kin}}{H_{\rm kin}'^2}\right)\approx 5
+ {\cal O} \left(\frac{h_{\rm kin}}{h_*}\right)^2 ~,
\ee
where, in the last step, we have again expanded Eq.~\eqref{Hdec} in the limit $h_{*}\gg h_{\rm kin}$. The prediction $f_{\rm NL}\sim 5$ is a characteristic signal of our mechanism in the limit of fast decay (i.e. when Eq.~\eqref{eq:modc} holds), and when $h_{*}\gg h_{\rm kin}$ so that ${\cal O} (h_{\rm kin}/{h_*})$ terms in Eq.~\eqref{Hdec} can be neglected. This analytical result is confirmed by the numerical analysis presented below. We remark that this level of non-Gaussianity is still compatible with the current bound $f_{\rm NL}^{\rm local} = -0.9 \pm 5.1$~\cite{Akrami:2019izv} at the $1\sigma$ confidence level. Importantly, the next generation surveys will reach the sensitivity required to either detect the signal or rule out the mechanism as the source of the primordial perturbation in the aforementioned limits. In particular, upcoming large-scale structure probes such as DESI and Euclid are expected to constrain the primordial non-Gaussianity with uncertainties of $\mathcal{O}(1)$~\cite{Font-Ribera:2013rwa, Amendola:2016saw, Mueller:2017pop}.

\noindent\textbf{Numerical results.}
The analytical approximations we proposed apply to a regime where the inflaton decays efficiently in the kinematically allowed window prior to the first zero crossing of the spectator field value. To go beyond this limit, we numerically solve the following system of equations  
\baq
\label{fluideom}
\dot\rho_{\phi}+3 H\rho_{\phi}  &=& -\Gamma(h)\rho_{\phi},
\\\nonumber
\ddot h+3 H\dot{h}+\lambda_h h^3 &=& 0,\\\nonumber
\dot{\rho_{\rm r}} + 4 H \rho_{\rm r} &=& \Gamma(h)\rho_{\phi},\\\nonumber
3 M_{\rm P}^2 H^2 &=& \rho_{\phi}+\rho_{\rm r}+\frac{1}{2}\dot{h}^2+\frac{\lambda_h}{4}h^4~, 
\eaq
where $\Gamma$ denotes the real part of Eq.~\eqref{eq:G}. The initial conditions are determined by matching the solution of the system composed by Eq.~\eqref{fluideom} with the first line replaced by $\ddot{\phi} + 3 H\dot{\phi} + V'(\phi) = 0$, at a matching time chosen well after the end of inflation (so that $\langle w_{\phi}\rangle = 0$) and well before $t_{\rm kin}$, so that $\Gamma = 0$. The initial conditions for this second system of equations are set by the inflationary slow roll solution, together with $\dot{h}_{*}=0, \rho_{r*} =0$ and the initial spectator field value $h_{*}$. Note that modelling the inflaton decay by inserting the diffusion term $\Gamma \dot{\phi}$ in its equation of motion would not be justified as $\dot{\phi}/\Gamma \gtrsim \phi$ when the decay takes place. We have checked that our results do not depend on the choice of the matching time within the given boundaries.

In our computation we set the inflaton parameters in Eq.~\eqref{eq:lag2} to $m_{\phi}= 1.55 \times 10^{9}$ GeV, $\lambda_{\phi} = 3.75\times 10^{-21}$ and $y_{\phi} = 1$, which yields $H_{*} \approx 1.6 \times 10^{10}$ GeV and $\epsilon_*\approx 0.017 \dots 0.018$ at the horizon crossing of the pivot scale $k_{*} = 0.05$~${\rm Mpc}^{-1}$. The precise values vary depending on the reheating process controlled by the spectator field couplings. For this choice of inflaton parameters, the inflaton contribution to the curvature perturbation is negligible, $\zeta_{\phi}^2 \sim 10^{-17}$, and will therefore be omitted henceforth. Note also that our results are not limited to the precise form of the inflaton potential used in Eq.~\eqref{eq:lag2}. Any other potential leading to same $H_{*}$ and $\epsilon_*$, and reducing to a quadratic form at the end of inflation, would give essentially the same results for the curvature perturbation induced by the modulation mechanism. 

\begin{figure}[t]
\centering
\includegraphics[width=1\linewidth]{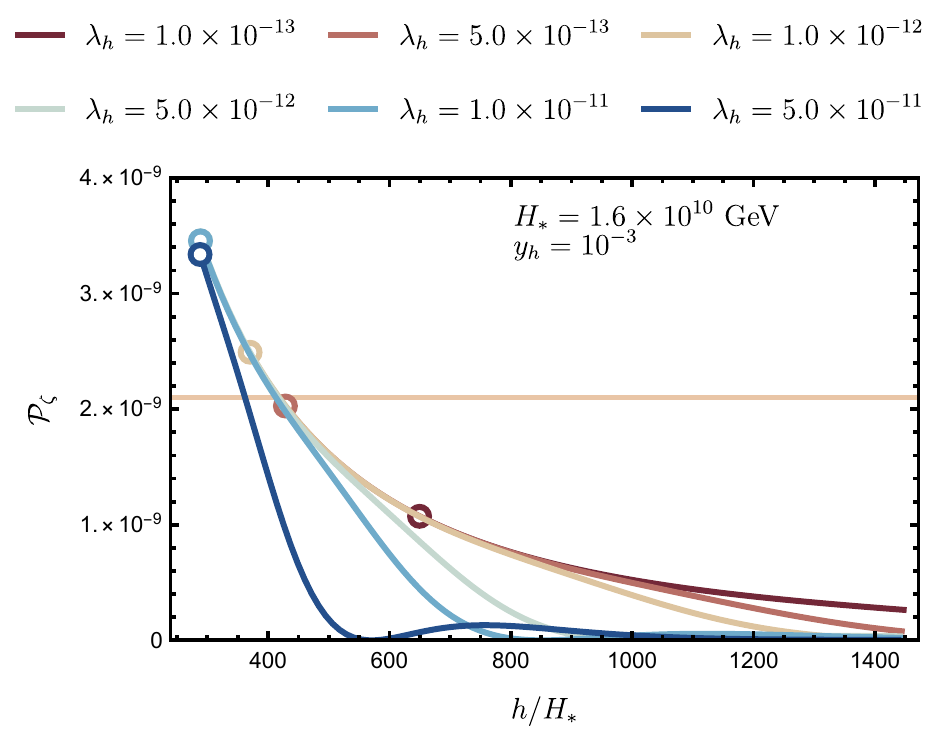}
\caption{\it The amplitude of the power spectrum sourced by the modulation mechanism computed at the pivot scale $k_* = 0.05\;{\rm Mpc}^{-1}$. Here $H_*$ denotes the Hubble rate at the horizon exit of $k_*$. The horizontal line indicates the central value of the Planck data \cite{Aghanim:2018eyx}. The circles mark the smallest field value which satisfies the mean-field condition assumed in our analysis. 
}
\label{fig:P}
\end{figure}

\begin{figure}[t]
\centering
\includegraphics[width=1\linewidth]{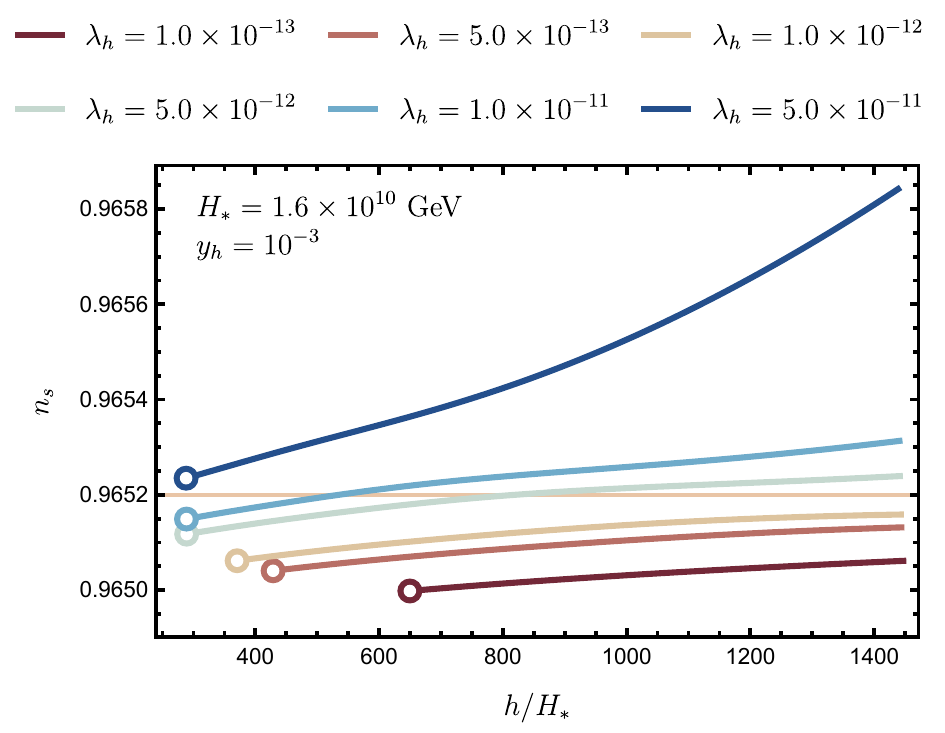}
\caption{\it The spectral index at the pivot scale $k_* = 0.05\;{\rm Mpc}^{-1}$. The horizontal line indicates the central value of the Planck data \cite{Aghanim:2018eyx}.
}
\label{fig:ns}
\end{figure}

\begin{figure}[t]
\centering
\includegraphics[width=1\linewidth]{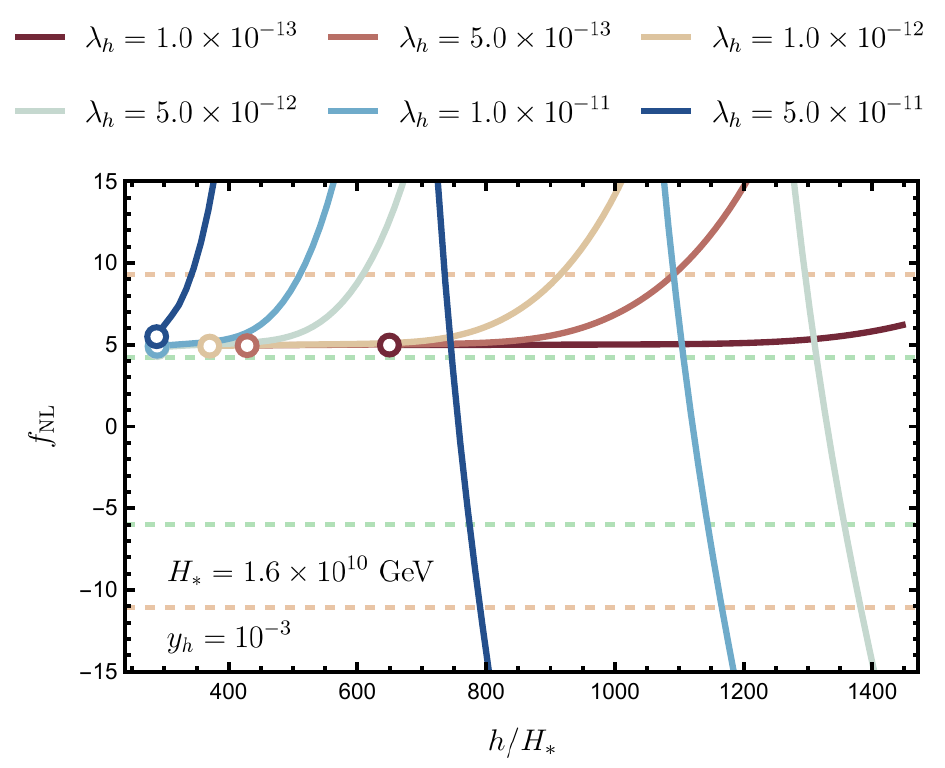}
\caption{\it The local non-Gaussianity parameter $f_{\rm NL}$ at the pivot scale $k_* = 0.05\;{\rm Mpc}^{-1}$. The green and orange horizontal dashed lines indicate the $1\sigma$ and $2\sigma$ confidence intervals of the Planck data \cite{Akrami:2019izv}, respectively. }
\label{fig:fnl}
\end{figure}

To compute the curvature perturbation, we construct a grid of initial values $h_*$ 
(the results shown in the figures are computed using a grid of $100$ points and a step size $\Delta \chi_* = 0.78 H_*$). For each grid point we numerically evolve the system Eq.~\eqref{fluideom} up to a common final $\rho$ chosen such that $\rho_{\phi}/\rho < 10^{-10}$ and determine the corresponding number of $e$-folds $N(h_*)$. 
From this data we numerically evaluate the first and second derivatives of $N(h_*)$ with respect to the initial field value $h_*$, corresponding to the coefficients of the $\delta h_*(k_*)$ powers in Eq.~\eqref{eq:deltaN}. For the spectator field perturbations at the horizon crossing $\delta h_*(k_*)$ in Eq.~\eqref{eq:deltaN}, we use the linear perturbation theory result for the two point function amplitude, ${\cal P}_{\delta h_*}(k_{*}) = (H_*/(2\pi))^2$, and neglect any non-Gaussianity in $\delta h_*(k_*)$. The spectrum of the curvature perturbation and the non-Gaussianity parameter $f_{\rm NL}$ are then given by 
\baq
{\cal P}_\zeta(k_*) &=& N'(h_*)^2 \left(\frac{H_*}{2\pi}\right)^2\\
f_{\rm NL}(k_*) &=& \frac{5}{6}\frac{N''(h_*)}{N'(h_*)^2}~,
\eaq
and the spectral index by Eq.~(\ref{eq:nsana}). All these quantities are readily evaluated from the numerical data. The results for ${\cal P}_\zeta$, $n_{\rm s}$ and $f_{\rm NL}$ at the pivot scale $k_* = 0.05\;{\rm Mpc}^{-1}$ are shown in Figs.~\ref{fig:P}, \ref{fig:ns} and \ref{fig:fnl}, respectively. We have checked that decreasing the final value of $\rho$ does not affect the results, confirming thereby that the curvature perturbation has relaxed to a constant well within the time span of the numerical computation.

The starting points of the lines in Figs.~\ref{fig:P}, \ref{fig:ns} and \ref{fig:fnl}, marked by circles, correspond to the smallest field value for which the mean field condition in Eq.~(\ref{meanfield}) holds for each value of $\lambda_h$. The fast decay region, where the inflaton decay essentially completes during the first zero crossing of the spectator field, is approximately determined by the condition Eq.~\eqref{eq:modc}, which for the parameters chosen as in the figures reads $h_*/H_* \lesssim  0.6 \lambda_h^{-1/4}$.
Increasing $\lambda_h$ makes the inflaton decay less efficient as $\Gamma(0)/H_{\rm kin}\propto (\sqrt{\lambda_h}/h_*)^{-1}$ and, correspondingly, the upper limit of the fast decay region decreases with larger values of $\lambda_h$. As expected, in the fast decay regime the numerical results agree relatively well with the analytical approximations in Eqs.~\eqref{eq:Pana}, \eqref{eq:nsana} and \eqref{eq:fnlana}. For the cases shown in Figs. \ref{fig:P} and \ref{fig:fnl}, the analytical estimates for ${\cal P}_{\zeta}$ and $f_{\rm NL}$ (obtained by using the full form of Eq.~\eqref{Hdec} in Eqs.~\eqref{eq:Pana} and \eqref{eq:fnlana}) respectively deviate less than $10$\% and $15$\% from the corresponding numerical results in the fast decay region $h_*/H_* \lesssim  0.6 \lambda_h^{-1/4}$.  

For $h_*$ values not in the fast decay region \eqref{eq:modc}, i.e. for $h_*/H_* \gtrsim  0.6 \lambda_h^{-1/4}$ when parameters are chosen as in Figs.~\ref{fig:P}, \ref{fig:ns} and \ref{fig:fnl}, the decay is no longer completed during the first window allowed by kinematics and the analytical estimates cease to be applicable.  For each value of $\lambda_h$ in Fig.~\ref{fig:fnl}, the end of the fast decay region $h_*/H_* \sim  0.6 \lambda_h^{-1/4}$ coincides with the regime where $f_{\rm NL}$ starts to grow towards large positive values.  Increasing $h_*$ further delays the inflaton decay successively to the second, third, or following zero crossing of the spectator field value, leading to a drastic amplification of the non-Gaussianity and increasingly complicated non-monotonous forms for both ${\cal P}_{\zeta}(h_*)$ and $f_{\rm NL}(h_*)$, as seen in Figs.~\ref{fig:P} and~\ref{fig:fnl}, respectively. This appears to be a generic feature of the setup, indicating that configurations not in the fast decay regime \eqref{eq:modc} tend to generate unacceptably large non-Gaussianity, possibly apart from tuned configurations around points where $f_{\rm NL}$ accidentally crosses zero.

\begin{figure}[h]
\centering
\includegraphics[width=1\linewidth]{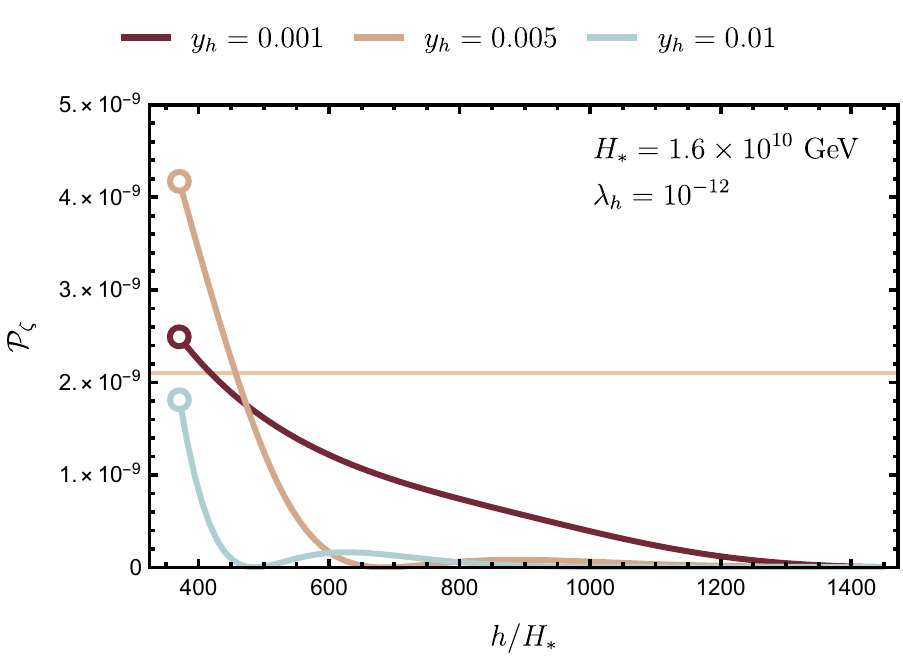}
\caption{\it The dependence of ${\cal P}_{\zeta}$ on the Yukawa coupling $y_{h}$. The horizontal line indicates the central value of the Planck data \cite{Aghanim:2018eyx} }
\label{fig:pvsy}
\end{figure}

\begin{figure}[h]
\centering
\includegraphics[width=1\linewidth]{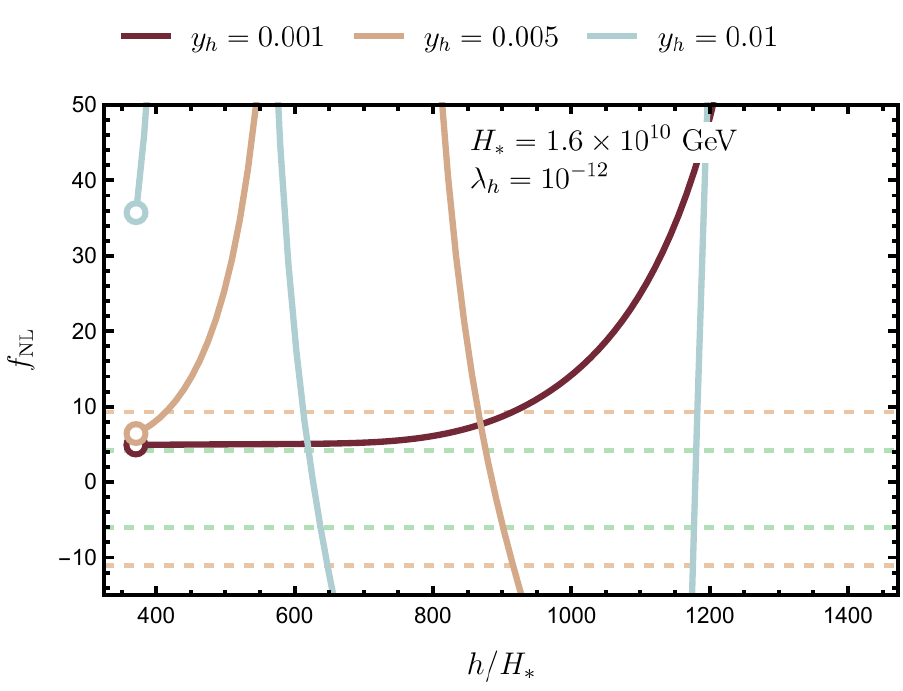}
\caption{\it The dependence of $f_{\rm NL}$ on the Yukawa coupling $y_{h}$. The green and orange horizontal dashed lines indicate the $1\sigma$ and $2\sigma$ confidence intervals of the Planck data \cite{Akrami:2019izv}, respectively.}
\label{fig:fnlvsy}
\end{figure}

As for the Yukawa coupling $y_{h}$, larger values of this quantity decrease $h_{\rm kin}$ in Eq. (\ref{hkindef}). This narrows the window where the inflaton decay is kinematically allowed around the spectator zero crossings and makes the decay process less efficient. Correspondingly, the region where the spectator field value crosses zero more than once before the decay, i.e. the region where the fast decay condition (\ref{eq:modc}) does not hold, is pushed towards smaller values of $h_*$ as $y_h$ grows. The dependence of ${\cal P}_{\zeta}$ and $f_{\rm NL}$ as functions of $y_{h}$ is illustrated in Figs. \ref{fig:pvsy} and \ref{fig:fnlvsy}, respectively, for a sample of $y_{h}$ values.

\noindent{\bf Implications for inflationary model building.} In our analysis, the inflaton potential affects directly only the value of the spectral index~\Eq{eq:nsana}. This is not specific to our setup and applies also to generic modulated reheating and curvaton models when the curvature perturbation is dominated by the spectator field. We remark that a generic prediction of such setups is a negligible tensor-to-scalar ratio $r$, which follows from the low inflationary scale required to keep the perturbations sourced by the inflaton subdominant. 
In particular, this means that potentials leading to too large tensor perturbations in the inflaton dominated limit can use the mechanism proposed here, as well as other modulated reheating or curvaton models, to achieve agreement with observations. An extensive analysis of inflaton potentials compatible with the curvaton framework was presented in {\it Encyclopaedia Curvatonis}~\cite{Vennin:2015vfa} and qualitatively similar conclusions apply to our setup. 

\noindent{\bf Embedding into the SM.} The above analyses relied on the interplay between the inflaton, a generic spectator field characterised by a quartic potential and a fermion that interacts with both the scalar fields. By regarding the parameters in the Lagrangian~\eqref{eq:lag2} as free and working at the level of classical potential, the presented analytic and numerical computations have highlighted the limits where the framework can match the CMB observations. In Ref.~\cite{Karam:2020skk} we have shown that the required modulation mechanism is already implemented in a well-known extension of the SM, where the particle content considers a singlet inflaton and right-handed Majorana neutrinos. It is therefore of interest to investigate whether the same setup allows for the generation of the CMB through the dynamics discussed in the present work. 

If the spectator field is to be identified with the Higgs boson, our analysis applies provided that the effective Higgs potential can be approximated by the tree level form:  $V(h) =\lambda_h(\mu)h^4/4$. The renormalization scale is set to $\mu\sim \sqrt{y_t} h$, with $y_{\rm t}$ being the top-quark Yukawa coupling, corresponding to the largest effective mass scale when $h_*\gg H_*$. The effective potential has a maximum where $\lambda_h(h_{\rm max}) = -\beta_h(h_{\rm max})/4$ and the presented analysis, which used $V = \lambda_h h^4/4$ with a constant coupling, is applicable only for $|h| \lesssim |h_{\rm max}|$. Using SM two-loop beta functions in the $R_\xi$ gauge and $\rm \overline{MS}$ scheme, for a (world average) top-quark mass of $m_t=172.9$ GeV we find $\lambda_h(h_{\rm max}) \gtrsim {\cal O}(10^{-5})$. However, to produce the observed spectral tilt in the quartic setup, we require $\lambda_h(h_*) < 10^{-8}$ -- see the discussion below Eq.~\eqref{eq:nsana}. Moreover, even smaller values in the ballpark of $\lambda_h(h_*) \lesssim  10^{-11}$ are necessary to avoid large non-Gaussianities, as shown in Fig.~\ref{fig:fnl}.
The specific points where $f_{\rm NL}(h_*)$ in Fig. \ref{fig:fnl} accidentally crosses zero might represent a possible caveat. However, although we have not studied this in detail, it is expected that even such configurations lead to too large non-Gaussianity in the form of the trispectrum amplitude $g_{\rm NL} \propto N'''/N'^3$. We thus conclude that, assuming the SM running of couplings, the Higgs boson seemingly fails to source the observed primordial perturbation through the mechanism studied here. 

One may of course ask if it is possible to circumvent these problems in extended setups with new physics coupled to the Higgs boson. Here we do not address this important question in detail, but we briefly comment on possibly viable phenomenological structures. 
First, a negative threshold correction $\lambda_h(h_*)\rightarrow \lambda_h(h_*) +\delta \lambda_h \lesssim 10^{-10}$ could make the scenario work, at the cost of fine-tuning  $\delta\lambda_h$ and $h_*$ close to the threshold scale. Also new scalar degrees coupled to the Higgs would act in favour of the scenario as they contribute positively to $\beta_h$ and make it possible to decrease $\lambda_h(h_{\rm max}) = -\beta_h(h_{\rm max})/4$. An interacting fixed point $\lambda_h(\mu)\lesssim 10^{-11}$, $\beta_h(\mu) = 0$ at, or below, the scale $h_*$ would also work, and it would be interesting to investigate if such fixed point could arise, for instance, from anomalously broken scale invariance. More generally, the tree-level Higgs potential could differ from the quartic form, for example due to the non-minimal curvature coupling $\xi R h^2$ or couplings to other fields with non-vanishing VEVs. This would change the reheating dynamics compared to our analysis, leading to potentially different conclusions. One could also think of scenarios where $h_*$ is located beyond the vacuum present at large field values, contrary to what we have assumed.  Suppose, for instance, that the corresponding minimum $h_{\rm min}$ is brought to sub-Planckian values by the interaction with an additional scalar field used to generate the right-handed neutrino Majorana masses via symmetry breaking. One could then arrange  $h_* > h_{\rm kin} > h_{\rm min}$, so that the reheating would commence at $h\sim h_{\rm kin}$ and, if thermal corrections rapidly lift the minimum $h_{\rm min}$, the Higgs could relax into the electroweak vacuum. A more careful assessment of these possibilities,  however, requires dedicated analyses that go beyond the scope of this work. 

\noindent{\bf Conclusions.} 
We have shown that a spectator field with a quartic potential can alone source the observed primordial perturbation through the modulated reheating mechanism, realised without direct couplings between the spectator and the inflaton field and by using only renormalisable operators. To obtain a 
red spectrum
, the spectator field needs to be sufficiently displaced from vacuum during inflation, its quartic coupling $\lambda_h \lesssim 10^{-8}$ and, as in generic spectator models, the inflationary dynamics needs to yield a suitable value for the slow roll parameter $\epsilon_{\rm H}$ at the horizon crossing of observable modes. The setup gives rise to primordial non-Gaussianities of the local type. We find that when the inflaton decay completes sufficiently fast (during the first spectator oscillation that lifts the kinematical blocking associated to the modulation channel), the bispectrum amplitude is set by $f_{\rm NL} \sim 5$, well in agreement with the current bounds and a testable signature by upcoming surveys. For $\lambda_h \lesssim 10^{-11}$, this is the case for a relatively broad range of initial spectator field values $h_*$. For larger values of $\lambda_h$, the inflaton decay is slower which gives rise to a growing tension between obtaining the observed amplitude of perturbations and maintaining the non-Gaussianity within the observational bounds.

Because the primordial perturbations are not directly related to the inflaton potential, the proposed framework, as well as modulated reheating and curvaton models in general, allows a wide range of inflationary models to come in agreement with present data, including those built on quadratic and quartic potentials. The predicted scalar-to-tensor ratio is generally negligible. 

The scenario we discussed allows, in principle, for a straightforward identification of the spectator field with the Higgs boson in a popular extension of the Standard Model with Majorana right-handed neutrinos. Although the required indirect modulation mechanism can be effectively implemented, matching the observed perturbation needs a modification of the Standard Model renormalisation group equations. In fact, using the Standard Model running and approximating the Higgs effective potential with the tree-level term, we find that the required value of the quartic coupling, $\lambda_h \lesssim 10^{-11}$, cannot be obtained in the regime where a quartic form well approximates the full potential. Whereas more precise studies that use the full form of the Higgs effective potential are needed in order to fully assess the possibility, we have briefly discussed promising phenomenological extensions of the Standard Model which could allow the Higgs boson to source the observed perturbations in the considered scenario based on a quartic potential. It would be of interest to study how these proposals could be realised in concrete particle physics setups.

\noindent\textbf{Acknowledgement.} This work was supported by 
 the EU Marie Curie grant 786564, the European Regional Development Fund through the CoE program grant TK133, the Mobilitas Pluss grants MOBTT5, MOBTT86, and the Estonian Research Council grants PRG356, PRG1055 and PRG803. 
A.R. was supported by the U.K. Science and Technology Facilities Council grants ST/P000762/1 and ST/T000791/1 and Institute for Particle Physics Phenomenology Associateship.
 
\bibliography{biblio.bib}
\end{document}